# *Is electricity storage systems in the Netherlands indispensable or doable? Testing electricity storage business models with exploratory agent-based modeling*

S.A.R. Mir Mohammadi Kooshknow [a]*, R. Herber [a], F. Ruzzenenti [a]

[a] *University of Groningen, Nijenborgh 6, 9747 AG, Groningen, the Netherlands*

**Abstract**

Electricity storage systems (ESS) are hailed by many scholars and practitioners as a key element of the future electricity systems and a key step toward the transition to renewables . Nonetheless, the global speed of ESS implementation is relatively slow, and among possible reasons is the lack of viable business models. We developed an agent-based model to simulate the behavior of ESS within the Dutch electricity market. We adopted an exploratory modeling analysis (EMA) approach to investigate the effects of two specific business models on the value of ESS from the perspective of both investors and the government under uncertainties in the ESS technical and economics characteristics, and uncertainties in market conditions and regulations. Our results show ESS is not profitable in most scenarios, and generally "wholesale arbitrage" business model leads to more profit than "reserve capacity". In addition, ESS economic and technical characteristics play more important roles in the value of ESS than market conditions, and carbon pricing.

**List of Abbreviations**

ESS, Electricity Storage System

ABM, Agent-based Modeling

EMA, Exploratory Modeling Analysis

EABMA, Exploratory Agent-based Modeling Analysis

RES, Renewable energy sources

CCGT, Combined-cycle Gas Turbine

OCGT, Open-cycle Gas Turbine

NPV, Net Present Value

CES, Community Energy Storage

TSO,  Transmission System Operator

# 1 Introduction

Electricity storage systems (ESS) have several potential ways for providing flexibility in the electricity system [1] and solving negative effects of variable renewable generation[2]. Despite some growth rate in implementation of ESS worldwide [3], finding viable business models for electricity storage systems has been a challenge for widespread implementation of ESS [4,5]. In a previous work [6], we identified barriers of ESS development, and accordingly developed a design space of ESS single-application business models for the Netherlands. Afterwards, in a second work [7], we provided a view on the testing process of business models, and we justified joint adoption of agent-based modeling (ABM) and exploratory modeling analysis (EMA) as our research approach to test and obtain insights on the benefits of ESS business models for various actors of the electricity sector.

Several agent-based models have been developed for analysis of problems related to energy transition and policy [8,9], including popular electricity market models such as PowerACE [10,11], EMLab [12,13], and AMIRIS [14]. The literature on use of ABM to address ESS problems can be generally classified into two groups of household- or community-scale, and grid-scale applications. In the former group, research projects incorporate the ESS's role in or study its effects on: demand response[15]; local electricity markets[16–18]; demand side management [19]; energy distribution among prosumers [20]; comparison of household and community ESS benefits criteria [21]; effects of household demands on the community energy storage [22]; and electric vehicle charging problem [23,24]. In this group, PV generation was part of the study in most cases [16–23]. In the group of grid-scale ESS applications, most studies employ the popular electricity market models. Genoese et. al. combined PowerACE and optimization models to compare benefits of flexibility offered by ESS and conventional power plants at the level of energy companies in the German electricity market[25]. Khan et. al. combines EMLab and optimization capabilities to investigate the effects of demand response and ESS on the need for capacity markets[26]. Fraunholz et. al. adopts PowerACE to test benefits of different settings of energy only markets and capacity remuneration mechanisms for electricity storage technologies [27]. Wehinger et. al. develops ABM to analyze effects of storage and electric vehicle clusters on the German electricity wholesale market using model predictive bidding for agents[28]. Lai et. al. integrates agent-based and power flow models for long-term electrical power system planning[29]. Size of scientific literature with explicit concentration on analysis of ESS business models using ABM is significantly small. Safarazi et. al. uses the AMIRIS model to simulate the profit and maximizing self-sufficient operation of CES business models [30]. Finally, to our knowledge, there is no previous research on stationary grid-scale ESS business models for the energy-only markets, with joint adoption of ABM and EMA methodologies.

In the current paper, we present detailed design and implementation of our exploratory agent-based modeling analysis (EABMA) approach to envisage the effects of grid-scale ESS business models on the investor's and the societal interests under deep uncertainties.

With this aim, we explore how and to what extent two ESS single-application business models[1] for the wholesale and balancing electricity markets ("wholesale arbitrage" and "reserve capacity") can serve the interests of the investors and the society under deep technical, economic, market, and regulatory uncertainties.

This paper is organized as follows: In Section 2, we explain details of design and implementation of the EABMA approach. First, we will identify the actors' decisions to investigate, the actors' goals and criteria to observe, and the external factors and uncertainties to take into account. Then, we will provide the details of the agent-based model design, implementation and validation. In section 3, we will elaborate on the extensive computational experiments and we will analyze them from various perspectives. Section 4 provides a note of the uncertainties and barriers for development of ESS with respect to the experiment results, and explains the limitations of the current study. Finally, in section 45, we will conclude with new insights and our future research.

# 2 Exploratory agent-based modeling Analysis

## 2.1 Analytical Framework

Building on previous work [7], the present analysis takes a further, decisive step into the EMA approach by developing four main elements: 1) means, interventions, and decisions of actors, 2) actor goals and evaluation criteria, 3) external factors and sources of uncertainty, and 4) an agent-based simulation model; for the simulation, we developed an agent-based model. Evaluation criteria are measurable variables with which the actors can evaluate to what extent their goals have been achieved. Means and interventions of actors are those actors' decisions that we are interested to analyze in this research. External factors are the variables which are assumed not to be influenced by the model. There is usually some degree of uncertainty involved in external variables, and most external variables in this research have deep uncertainty (which means that a probability distribution cannot be defined).

Actors in this study can be classified into commercial and government actors. Commercial actors include companies and end-consumers of electricity who are legally allowed to buy and sell electricity (in the Dutch regulatory framework). Government actors include all bodies which are not allowed to trade electricity, and their activities include law making, enforcement of the law, managing the markets, and maintaining the power grids. Commercial actors seek to maximize their profit in a sustainable way. Net present value is one criterion to measure profitability. In contrast, the government seeks to serve all sections of the society by ensuring availability of electricity (security of supply), affordability of electricity (fair prices for producers and consumers), and environmental acceptability of electricity. In the electricity system, the number of blackouts over time is a criterion

[1] In the single-application business model the value proposition includes only one of the ESS applications. (a map of potential ESS single-application business models for a European electricity market can be found in our previous work [6])

for measuring reliability, electricity price is a criterion for measuring affordability, and volume of total $CO_2$ emission is a criterion for measuring environmental acceptability of the system.

The main decision to consider in this study is the selection of business models by energy companies for storage of electricity. The companies make other decisions such as investments and bidding/offering in the electricity markets. We do not directly consider the decisions of the government and all its related bodies in this study, but they are indirectly considered as they influence some external factors in the study. External factors in the study include information on availability of renewable sources, price of fuels, price of $CO_2$, electricity demand, growth rate of renewable and non-renewable share in the electricity portfolio of the country, costs of energy storage systems(EES), and desirability of ESS.

The current study is limited to two grid-scale and single-application ESS business models: "Wholesale arbitrage" and "reserve capacity". In the wholesale arbitrage business model, an energy company buy/sell electricity from/to another energy company via the wholesale electricity market, and the profit depends on the price gap in the wholesale market and ESS costs. In the reserve capacity business model, an energy company buys electricity in the wholesale markets from other energy companies and sells in the balancing energy market to TSO (Transmission System Operator), and the profit depends of the difference of prices between the two markets and the ESS costs[2] (a more extensive explanation of the business models of concern can be found in our previous work [6]).

The analytical framework of the current research is illustrated in Figure 1. In this study, the agent-based model encodes the target system which is composed of electricity markets and storage options. The system is influenced by the decisions of actors and external factors, and the changes in the system are measured via four evaluation criteria variables.

[2] Note that balancing service can be in forms of capacity or energy. Balancing capacity providers in the Netherlands participate in yearly auctions and their balancing energy prices are determined with pay-as-bid model. Offering balancing energy is also possible for other participants and their balancing energy price will be determined with marginal pricing. Despite the term "capacity" in the "reserve capacity" business model, in the current paper we assume that ESS only provides balancing energy.

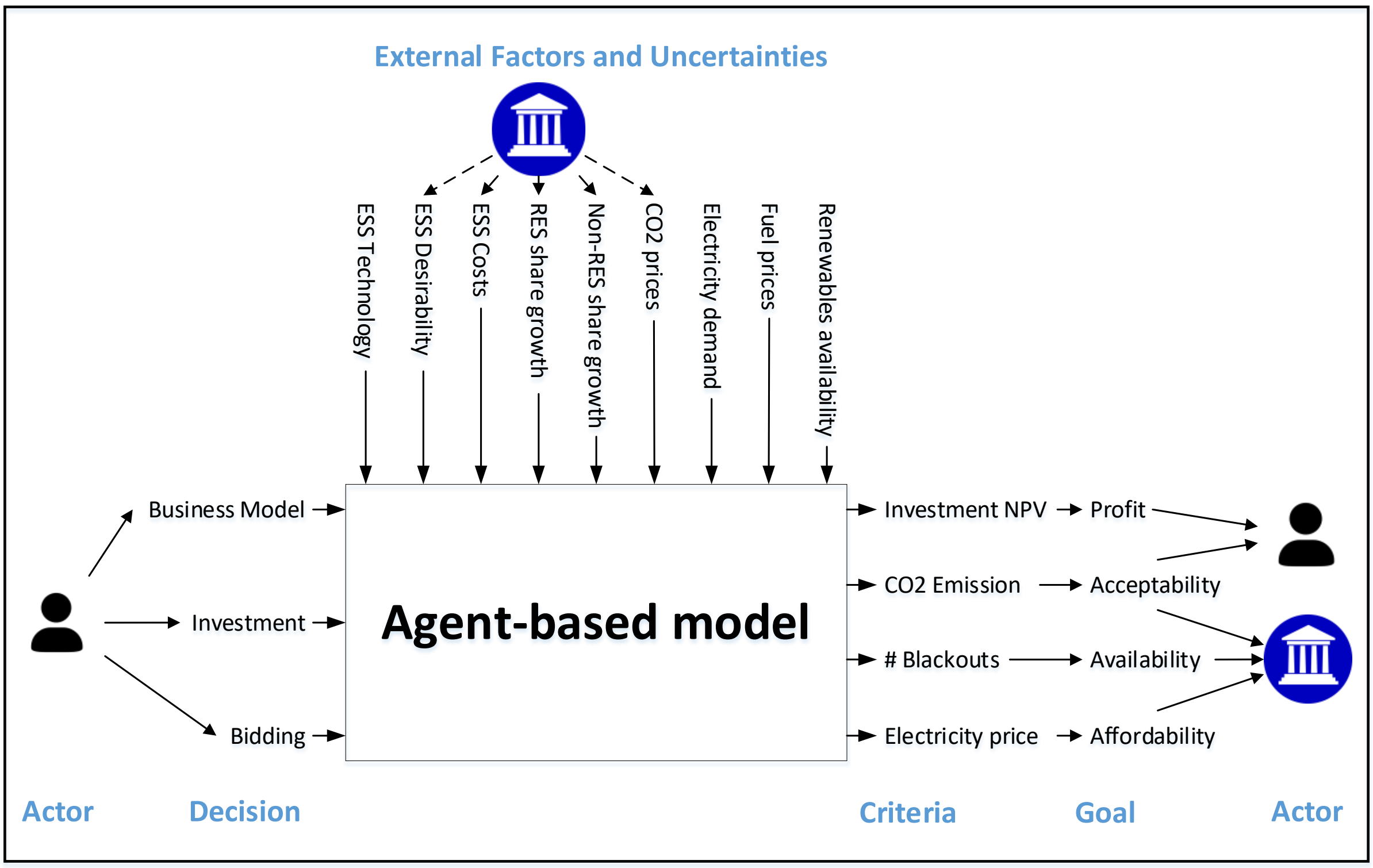

*Figure 1 Analytical Framework: The agent-based model encodes the target system which is influenced by the decisions of actors and external factors. The evaluation criteria are system variables which indicate to what extent the goals of actors are met.*

## 2.2 Agent-based Model

### 2.2.1 Model Design

#### *2.2.1.1 Model Scope*

Designing a model to explore ESS business models starts from designing the space in which the products and services of ESS can be traded. In this step, we start with modeling of electricity day-ahead and balancing markets for the Netherlands. For simplicity, we assume that these two markets represent all energy trading within the country, therefore other markets such as long-term contracts and intra-day markets are not included in the model. It can be justified because the price of long-term contracts (in forward and future markets) converge to spot prices, and the size and liquidity of intra-day markets is negligible. In addition, the model is assumed to be a single-country model, and there is no import and export of electricity, and $CO_2$ price is incorporated in the model and as an external variable.

#### *2.2.1.2 Agents, physical entities, and interactions*

Main agents in the market model are large producers, large consumers, retailers, and one market operator which represents the government in the model. These agents form the institutional network of the model and they interact with each other through information and cash links. Besides, these decision-making agents may own, control, or contractually represent some physical entities such as a grid, power plants, ESS, or loads[3]. For this paper, the ownership of ESS is limited to energy producing companies. The physical entities form a physical network, in which the entities are connected to a single grid[4] via wire links and the electricity is transported from supply points to demand points through the wires and the grid. The connection between the institutional network and the physical network is made via ownership and supply contract links. Decomposition of the model is illustrated in Figure 2.

All agents and physical properties in the model have attributes and behaviors. For agents, one group of behaviors is decision making. The most important decisions that we are interested in this research are the decisions for the investment on and the business models of ESS. Further details of attributes, decisions and behaviors of agents and physical properties are provided in Appendix A, respectively. In addition, details of bidding strategies of market participant are provided in 1.1.1.1.1.1Appendix CAppendix C.

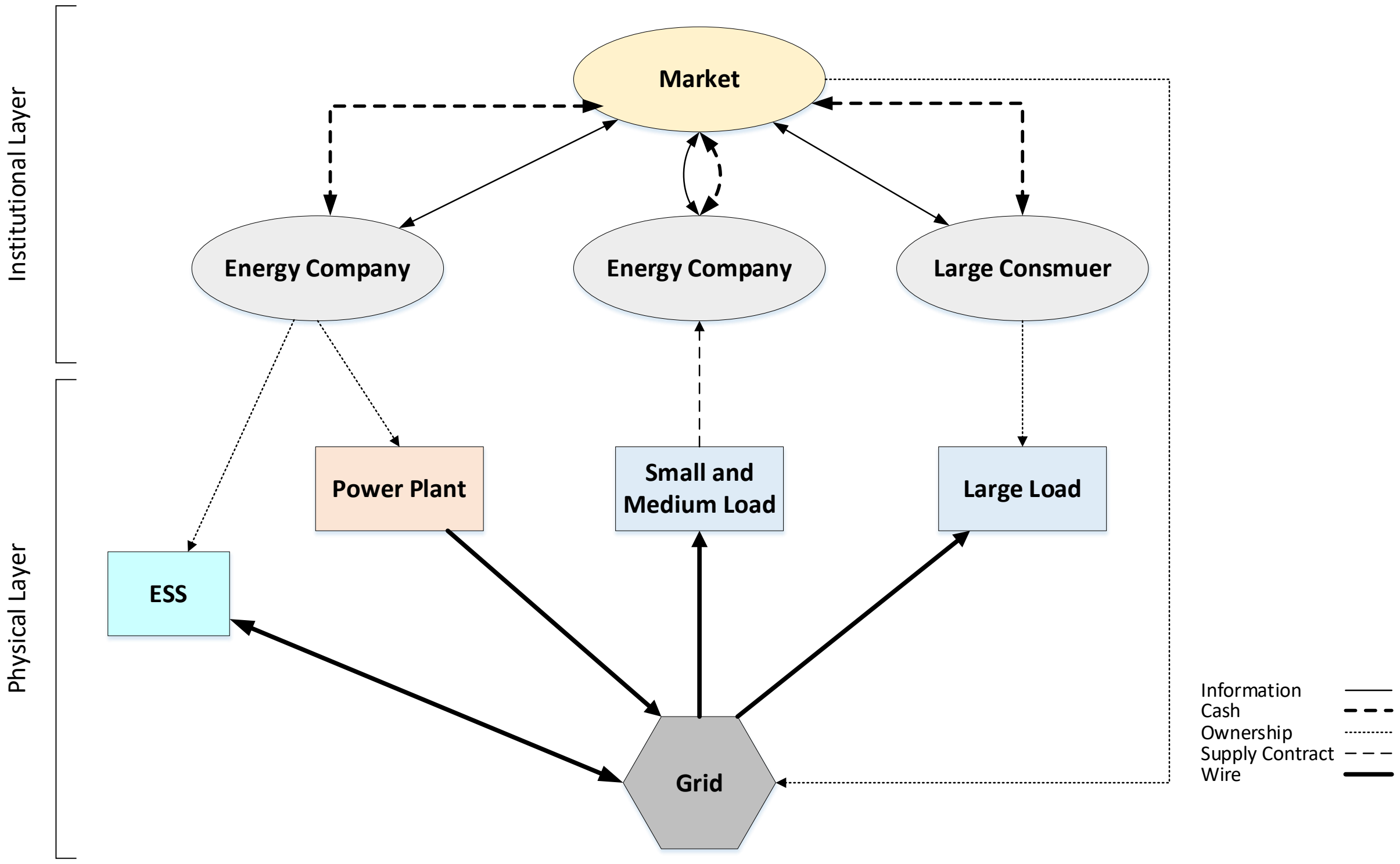

*Figure 2 Agents, physical entities, and the types of interactions among them in the agent-based model*

### 2.2.1.3 Environment

The environment of the model contains networks in which agents are connected to each other through different types of links. In addition, a number of parameters and variables are incorporated

[3] In this model, large loads refers to the loads connected to the high-voltage transmission grid, while other loads are connected to the distribution grid.
[4] In the current version of the model, the grid is modeled as a single node and no grid topology is taken into account

in the environment. In this study Model parameters are the modeler's assumptions about the agents and physical entities. The environment parameters and variables are provided in Appendix A.

#### *2.2.1.4 Model process*

The operation of the model starts with one-time initialization in which all agents and environment are built and their initial attributes are set. Then, the recurrent operation occurs as follows:

Every year, the generation, load, and ESS fleets change (new units are added or some current units are removed) based on the relevant scenarios' growth rates. Then, every hour, the availability of physical generating/consuming entities (power plants, loads, ESS) are determined based on their reliability value. Then they receive the data of fuel prices, wind condition, sun condition, and load profiles from the environment. They process the data, calculate/report available capacities, marginal costs, and willingness-to-pay, and send the data to their owners/controllers who participate in the markets as seller or buyer. All electricity sellers offer their electricity in the day-ahead market based on the marginal cost[5] , and the electricity buyers bid to the market based on their maximum willingness-to-pay (see calculation details of bidding strategies in Appendix CAppendix C). Large producers and consumers also offer and bid in the upward and downward energy for the balancing market. The market agent collects all the bids and offers, clears a double-auction wholesale market, informs the environment about the market price and volume, and informs market participants about their approved electricity programs. Based on the programs, the physical entities owned/controlled by the wholesale market participants inject electricity in or withdraw it from the grid. The grid physical entity sends a signal to the market agent if there is any imbalance in the wired network. Any real-time deviation from the electricity programs results in the activation of a balancing mechanism in which balancing bids and offers are processed by the market agent. The market agent determines balancing price, volume and direction of balancing and reports them to the environment. It also send the balancing programs to the market participant who will adjust or curtail generation/consumption of their fleet accordingly. The market agent also mediates the transfer of money from electricity buyers to electricity sellers. In addition, the market agent measures the CO2 emission from the physical properties of agents and charges them with the CO2 price. At the end of every hour, ESS entities update information about their content and NPV, and they report it to the environment.

The recurrent operation of the model is depicted in Appendix D inspired from UML sequence diagrams.

### 2.2.2 Implementation

The model is implemented in NetLogo [31] environment. It is a free and open-source software which enables rapid model development. The model is implemented such that it enables us to study various dynamics both at a scale of hours and a scale of years. The smallest time step in the model is represented by one hour. To manage computational intensity of the model as well as incorporating the seasonal patterns in the analysis, we assumed every 24 hours represent a month, and every twelve months represents a year. Therefore, a year consists of 288 time steps in the model. Pseudo code of the model is provided in Appendix EAppendix E .

[5] The cost of producing/offering additional electricity. This is the minimum rational price that a producer can offer in the market to compete with others.

### 2.2.3 Verification and validation

The model is verified using different methods such as recording and tracking agents behavior, single agent testing, and interaction testing in minimal mode (further details about the employed verification methods can be found in [32]).

We take an exploratory modeling analysis approach to manage the deep uncertainty in the model factors instead of consolidative modeling, in which a model is developed by consolidation of all detailed facts in order to produce estimates close to real-world values, (further explanation on the differences between the approaches can be found in [33]). Therefore, our aim is to find patterns and trends in the system. Exploratory modelling analysis produces knowledge even where strict model validation is not possible [34]. Nonetheless, we devote efforts for proper validation of the model by comparing the model results with the real world data. Because the size of existing ESS capacityin the Dutch electricity market was negligible, the validation process mainly focuses on the electricity market itself as the market is the environment in which ESS business can evolve. In this study, ESS is considered as a sort of combined power plant and load, therefore these functions for ESS can be verified and validated in parallel with other electricity generating or consuming agents. For validation of market behavior, we confront the electricity price in the model with the historical data, and we tune some parameters (e.g. maximum willingness to pay of consumers) in order to produce the nearest output to the real world data. We use historical data for the period 2016-2018 in the validation process (details of model assumptions and input data can be found in Appendix B). Figure 3 illustrates the development of simulated electricity wholesale price in the model and its correlation with the real-world data.

(a)

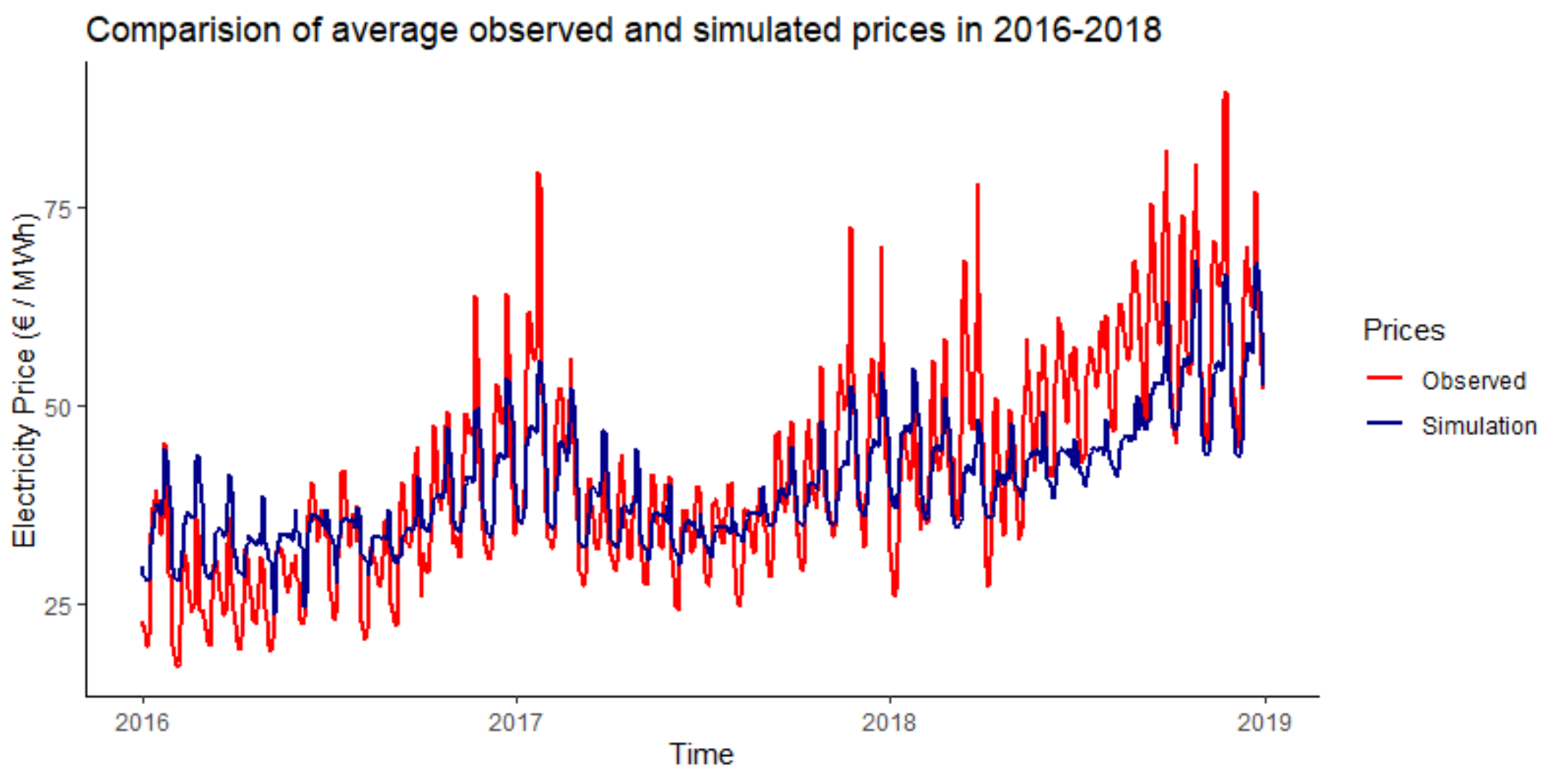

(b)

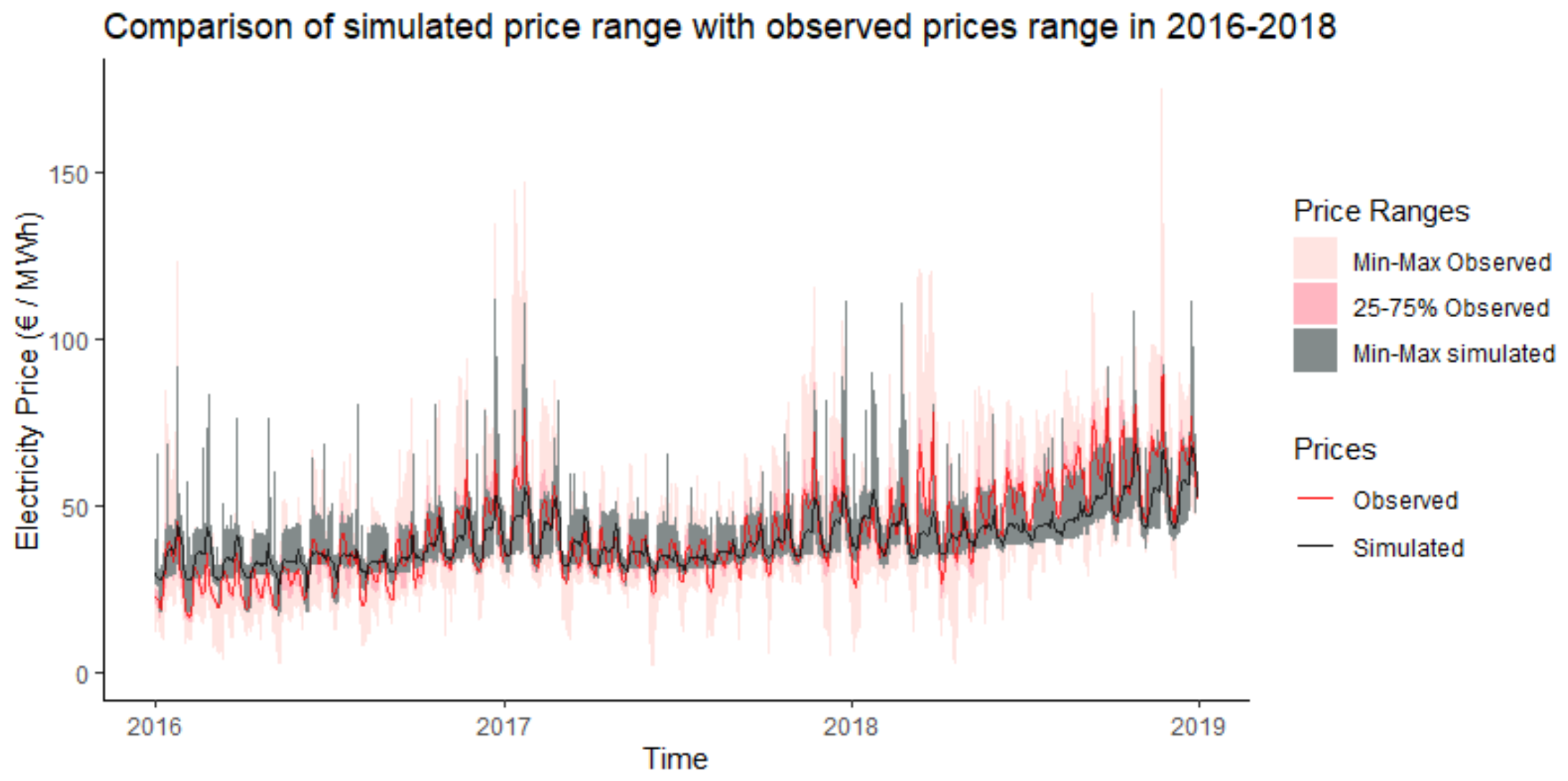

*Figure 3 Comparison of observed and simulated electricity prices: (a) comparison of averages (b) comparison of simulated price range with observed price range | In (a) we observe how a trend of average prices is reproduced by the model between 2016 and 2018, whereas in (b) we notice that range of simulated prices falls within the range of real observed price (Min-Max Observed) and it mostly overlaps with the range between first and third quartile of prices (24-75% observed)*

Because of the model assumptions (e.g. assumption that all electricity is traded in the spot market, assumptions on consumers' willingness to pay, etc.), as expected, the simulated prices are not completely aligned with the real-world prices, but the trend and most dynamics are reproduced which provides a good basis for the exploratory modeling analysis.

# 3 Exploring effects of ESS Business models

## 3.1 Design of Experiments

To investigate the effects of business models and functioning of ESS on the behavior and goal of the different actors involved (see Figure 1), we assess number of evaluation criteria variables under different scenarios. These variables are:

- Average NPV of ESS projects [€]
- Electricity wholesale price [€/MWh]
- Cumulative $CO_2$ Emission [ton $CO_2$]
- Cumulative blackouts hours (over 20 years) [Hours]

Detailed calculation of the evaluation criteria is provided in Appendix FAppendix F . The experiment scenarios are determined by combining different values for model parameters. Because of NetLogo limitations on the number of scenarios, we considered some extreme values for most parameters. Therefore, for some parameters we considered only two values as maximum and minimum, and for a couple of parameters three values are considered including a medium value. The parameters' values which form experiment scenarios are illustrated in Table 1. For other parameters a single value derived from literature, historical data, or model calibration is used (see Appendix B).

*Table 1 Value of the parameters in the experiment scenarios*

| Parameters | Values |
|---|---|
| ESS Business Model | Wholesale Arbitrage , Reserve Capacity |
| Desirability of ESS for Energy Companies (%) | 0,50,100 |
| Grid ESS capacity (MW) | 10,1000 |
| Maximum ESS project energy rating (MWh) | 10,1000 |
| ESS power capital cost (k.€/MW) | 1,100 |
| ESS energy capital cost (k.€/MWh) | 1,100 |
| ESS roundtrip efficiency (%) | 70,85,100 |
| Annual growth rate for RES (%/y) | 0,25 |
| Annual growth rate for non-RES (%/y) | -10,0,10 |
| Growth rate of $CO_2$ price (%/y) | 0,10 |
| Annual Growth Rate of Total Electricity Demand (%/y) | 0,2,4 |

The selection of values for scenario parameters was based on the technical and economic characteristics of ES technologies presented in [6] and the historical data on growth of generation capacity and demand. The experiments are designed to compare two single-application ESS business models: wholesale arbitrage and reserve capacity (for overview of ESS single-application business model please see [6]).

- For desirability of ESS we considered three scenarios in which either none, or half, or all of the energy producing companies are interested to add ESS to their fleet. The first simply represents the no ESS scenario.
- For grid capacity, two scenarios of 10 and 1000 MW are considered where the former more or less represents the status quo and the latter can be considered as an extreme value in the context of the Netherlands (given other existing flexibility providers such as gas-fired power plants).
- For energy rating, we set a maximum for individual ESS "projects". The two number present two orders of magnitude instead of presenting two exact numbers. For many electricity storage technologies an energy rating of 1 MWh is currently achievable, whereas 1000 MWh is currently existing with pumped hydro and compressed air energy storage system. Therefore, the two values can be considered as proper extremes for new ES technologies (see more information of energy rating of ES technologies in [6]).
- For capital cost of power and energy units we considered an upper value of 100 k.€/MW and 100 k.€/MWh, respectively. This is the current order of magnitude of capital costs for most ES technologies (see details of ESS capital costs in [6]). As an extreme for reduction of capital cost of the power unit and energy unit we considered 1 k.€/MW and 1 k.€/MWh, respectively.
- Round-trip efficiency of most ES technologies is currently between 60% and 90%. In the experiments we considered the efficiencies 70%, 85%, and 100% as normal, high and extremely high scenarios. Of course, an efficiency of 100% is impossible, but it is included in the analysis to explore the economic behavior of ESS when the efficiency goes above 90%.

- Currently, RES capacity in the Netherlands is less than the national plans. We basically considered two extreme growth rates. One extreme is zero growth and the other extreme in 25% annual growth which can make the electricity generation in the country completely RES-based in about 20 years.
- For non-RES capacity in the Netherlands we observed a negative growth rate for coal (~-12%) and almost zero growth for gas and nuclear between 2016-2018. We explore growth rates -10%, 0%, and 10% for non-RES capacity. The rate -10% means that within twenty years the non-RES capacity becomes about one-tenth of the current capacity, the 0% rates keep the capacity as it is now, and the 10% rate increases the capacity six times within twenty years of the experiment. The rate of 10% growth is not realistic, but we included it in our analysis to explore such a hypothetical future.
- The historic growth rate of $CO_2$ price in the Netherlands between 2016 and 2018 was 72%, however in the years to follow we considered two growth rate scenarios of 0% and 10% for this variable.
- The demand growth in the Netherlands between 2016 and 2018 was about 1%. We included two scenarios with 0% and 2% to cover the actual growth, and we included 4% growth to explore a more rapid electrification scenario.

The combination of all parameters in Table 1 generates 10368 scenarios. To capture variability and randomness, for each scenario we run the model 20 times[6] which results in 207360 experiments. In each experiment, we run the model for a period of 20 years (or 5760 model hours) which is proportional to the lifetime of most ESS technologies. The experiments were executed on 40 cores (2.6 GHz) of a single node in the Peregrine computer cluster and it was completed after approximately 142 hours.

## 3.2 Analysis of Results

To find patterns in the results of the experiments we defined four goal variables: 1) *profitability*; 2) *affordability*, 3) *acceptability*, and 4) *availability*; expressed by min-max normalization of the observed values of 'average NPV of ESS projects'; 'average wholesale electricity price'; 'cumulative $CO_2$ emission' and 'cumulative blackouts' respectively. With min-max normalization, all the values were scaled between 0 and 100. In addition, another variable *government goal* is defined as a weighted average of *affordability*, *acceptability*, and *availability* values. For simplicity, in this analysis, we considered equal weights for the three variables to calculate the *government goal*.

### 3.2.1 A big picture

Figure 4 illustrates the scatter plot of the relationship between the degree of ESS profitability in 10368 scenarios and the degree to which the government goals in those scenarios are met. In addition, the plot illustrated a profitability threshold line (the dashed line) that indicates a degree of profitability in which NPV = 0. All the scenario points below the line have negative average NPV and all the points above the line have positive average NPV. The color of points in the plot indicates whether the scenarios are absolutely profitable or not. Scenarios are absolutely profitable if all their 20 simulation instances have positive average NPV for ESS projects. In Figure 4, we observe some points are above the threshold line and have positive average ESS NPV, but their color indicates that they are not absolutely profitable. In such scenarios at least one out of twenty simulation instances led to negative NPV for ESS. Therefore, there is more risk of economic loss in such scenarios.

[6] This an arbitrary number selected by the modeler. In the experiment, running simulations less could not capture the variability of results well, and running the simulations more did not add value to the analysis.

In Figure 4, we observe that the majority of scenario points in the experiments fall below the threshold of profitability. We also observe that a considerable number of scenario points above the line are not absolutely profitable and their profit is subject to higher risk. In addition, we observe that the points are not uniformly distributed horizontally or vertically. Especially with respect to the government goals, we observe clear breaks in the space. To obtain deeper insight of such distribution, we continue with the analysis ESS profitability for energy companies.

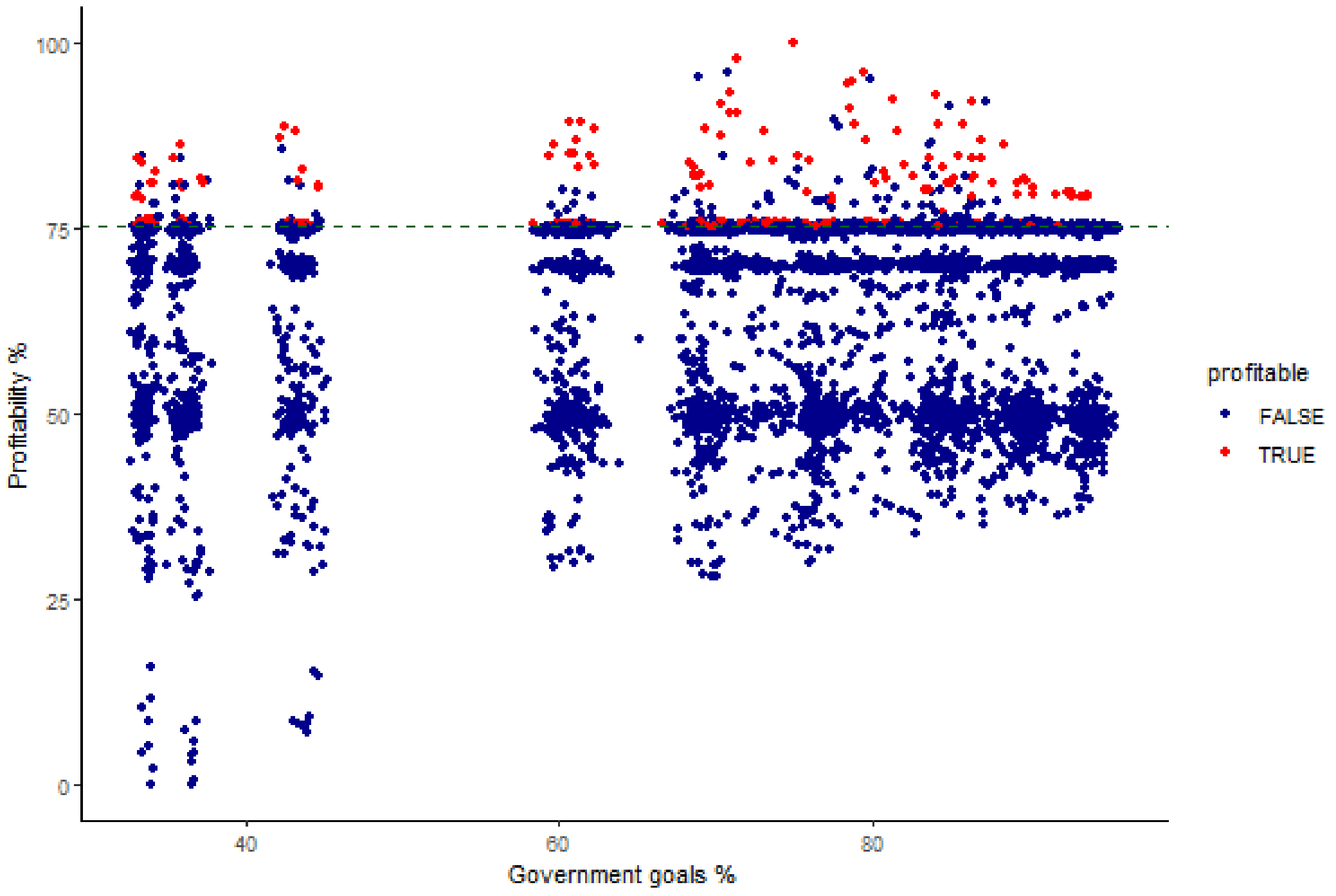

*Figure 4 Relationship between the degree of ESS profitability and the degree government goals are met. Red points are the scenarios in which all simulation runs have positive average NPV for ESS projects. Blue points indicate scenarios in which at least one of the simulation runs has negative average NPV for ESS projects.*

### 3.2.2 ESS Profitability

The experiment results contain extensive data for values of several parameters and the target variable (profitability). In order to reveal the hidden patterns in the data and find the most influential parameters in the ESS profitability, we employed a supervised machine learning technique. Because of the continuous nature of our variables, we adopted a regression tree method in the statistical software package “R”. Regression trees are mainly adopted for forecasting target variables in different segments of the parameter space. They also indicate which variables can significantly split the space for prediction of the target variable.

We found eight variables as important variables to split the space and predict the ESS profitability. The eight variables are ESS energy capital cost, maximum ESS energy rating, grid ESS capacity, ESS round trip efficiency, ESS Business Model, annual growth rate for non-RES, ESS power capital cost, and annual growth rate for RES.

We find that in most scenarios, ESS is not profitable, meaning that in most scenarios, the revenues are not sufficient to cover the capital and operational costs. In addition, in several scenarios, even profitable ones, the average degree of profitability is close to the breakeven point (NPV=0). Such patterns can explain concerns of investors for development of ESS.

Among profitable situations, the wholesale arbitrage business model leads to more profit than reserve capacity. It can be justified by the difference in the size of the day-ahead and balancing markets. The former provides more space for ESS to sell its stored electricity and earn revenue. Furthermore, it is noteworthy that reverse capacity is only profitable amid negative non-RES growth rate (fossil divestment), where controlled generation is replaced by variable RES generation.

We observe an explicit lack of profitability, as intuitively expected, in cases with very high ESS energy capital cost and very high ESS energy rating (which means bigger size and thus very high capital costs). In addition, we generally observe a lack of profitability in cases with high ESS power capacity in the grid and very high ESS power capital cost (which in turns leads to high capital costs again). This pattern has an exception where the ESS energy rating is high, ESS energy capital cost is low, and the round-trip efficiency is very high. In this condition, we observe some profitability potentials for ESS. It is an indication that a cheap ESS energy unit can cancel out the effects of an expensive ESS power unit if ESS can utilize most of its potentials, but the opposite is not true[7].

Most profitable scenarios for both business models are a result of a low ESS energy capital cost. In this condition, more profitable scenarios can be found in combination of low energy rating and low ESS power capacity, or of high energy rating and high ESS power capacity in the grid. In the former, the frequency of occurrence of profit is higher, and in the latter the value of NPV is higher; meaning that scaling up ESS leads to a lower probability of positive NPV, but it widens the profit margin.

Among profitable scenarios, there is a general pattern that negative non-RES growth rate and moderate RES growth lead to more profit. These patterns can be justified by knowing that lack of controlled Non-RES generation will escalate the problems of intermittent RES generation and ESS can play an important role to solve this problem. The problem of intermittent RES generation may increase with a moderate RES capacity growth rate. But, in case of very high RES capacity growth, there is abundance of supply which automatically solves the problem of intermittency in RES generation and reduces the margin of ESS in both the arbitrage and balancing markets significantly.

[7] It is worth reminding that the capacity of ESS is two dimensional, defined by both the speed at which it can be charged/discharged (power rating) and the amount of energy that can be stored (energy rating).

In addition, very high round-trip efficiency is a must for profitability of ESS. We observe no absolute profitability in the experiment instances for low round-trip efficiency. Last but not least, in some situations , we observe that despite a high value of average NPV, none of the internal scenarios is absolutely profitable and more degree of risk for that profit should be considered for them.

An illustration of the degree of ESS profitability with respect to the scenarios formed by the eight aforementioned parameters is provided in the form of a heat map in Figure 5. Rows and columns of the heat map show the different values for eight most influential scenario variables (ESS energy capital cost, maximum ESS energy rating, grid ESS capacity, ESS round trip efficiency, ESS Business Model, annual growth rate for non-RES, ESS power capital cost, and annual growth rate for RES). The heat map consists of several "tiles" at intersections of rows and columns, and each tile represents a scenario defined by its row and column. The color of the tiles shows the degree of ESS profitability in that specific scenario (red=profit, white= zero profit, blue=loss). In addition, each tile contains the information of four less influential variables which in total form 36 internal scenarios. Dots and their sizes reveal information about profitability of ESS in the internal scenarios. Dominance of the white color indicates that in most scenarios, the average NPV of ESS is close to 0. We observe more red color and more dots in the upper half of the maps which indicates more probability and value for profitability of Wholesale arbitrage (scenario "A" in the heat map). In addition, a small population of dots indicates lack of ESS absolute profitability in most scenarios. Besides, more red color and bigger dots appears in scenarios in which energy capital cost is low and the size of power and energy units are both low or right simultaneously.

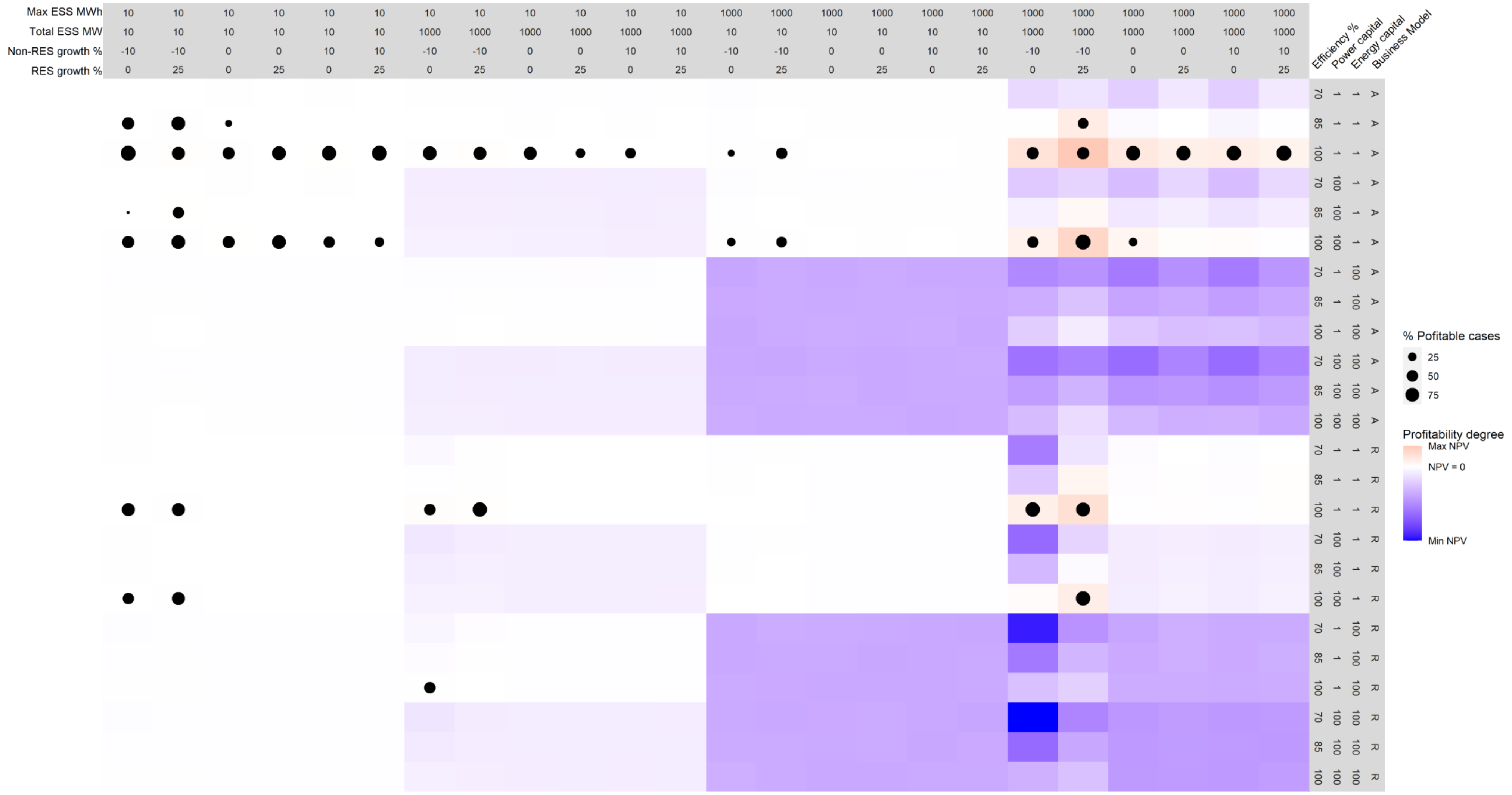

*Figure 5 Heat map of ESS profitability in different scenarios| Business Models: A = Wholesale Arbitrage, R = Reserve Capacity | Power Capital cost in €/MW, Energy Capital Cost in €/MWh| The heat map consists of several 'tiles' and each tile represents a scenario formed by a combination of values of the eight important parameters. In this plot, the color of each tile shows the degree of profitability of ESS in that tile. Each tile contains information of 18 internal scenarios that are formed by the four least important parameters. To reveal more information per tile, dot sizes are used to show how many percent of internal scenarios in each tile are absolutely profitable (= all simulation replication results in positive average NPV for ESS).| Dominance of the white color indicates that in most scenarios, average NPV of ESS is close to 0. We observe more red color and more dots in the upper half of the maps which indicates more probability and value for profitability of Wholesale arbitrage. | Small population of dots indicates lack of ESS profitability in most scenarios. |More red color and bigger dots appear in scenarios in which energy capital cost is low and size of power and energy capacity are both low or right at the same time.*

### 3.2.3 Government view

Adopting the regression tree method, we found that only three parameters were significantly influential to predict the values of the government goal score (= weighted average of availability, affordability, and acceptability scores). These parameters are “annual growth rate for RES”, “annual growth rate for non-RES”, and the “annual growth rate of total electricity demand”. In other words, goals of the government are mostly dependent on supply, demand, and generation sources, regardless of ESS presence and characteristics.

Predictably, we observe that higher RES growth rates, as much as lower demand growth rates, result in higher government goal scores. Significantly, negative non-RES capacity growth which results in phase-out of non-RES technologies leads to lower government goal scores. This is the main explanation of horizontal breaks in the space of Figure 4.

As mentioned before, the government goal score is the average of degrees of affordability, acceptability, and availability. The degree of affordability is measured by the electricity price which depends on the supply and demand. Any positive growth in the supply (RES or non-RES) results in lower electricity price and higher affordability, while any positive growth in the demand leads to higher price and lower affordability of electricity.

Furthermore, we observe that the highest acceptability emerges in a case of negative non-RES capacity growth. This is equal to rapid phase-out of non-RES and it results in less $CO_2$ emission and increase of environmental acceptability of the electricity. In case of zero or positive non-RES growth, we observe that higher growth rates in RES capacity results in higher acceptability degrees because of less $CO_2$ emission. In addition, more demand growth results in less acceptability, because it necessitates more generation of electricity which will be partially done by non-RES capacity, and it results in more $CO_2$ emission.

Finally we observe that in the presence of non-RES capacity, the availability of electricity is very high, which means that the number of blackouts is very low. In addition, in the absence of non-RES capacity (as a result of negative non-RES growth rate), more RES capacity growth rate results in more availability of electricity, and more demand growth results in less availability of electricity and is subject to more blackouts

More details about the distribution of governmental goals with respect to the influential parameters (annual RES growth rate, annual non-RES growth rate, annual demand growth rate) can be observed in Figure 6. In this figure, in addition to the government’s goal score, the changes in distribution of electricity affordability, acceptability and availability scores can be observed.

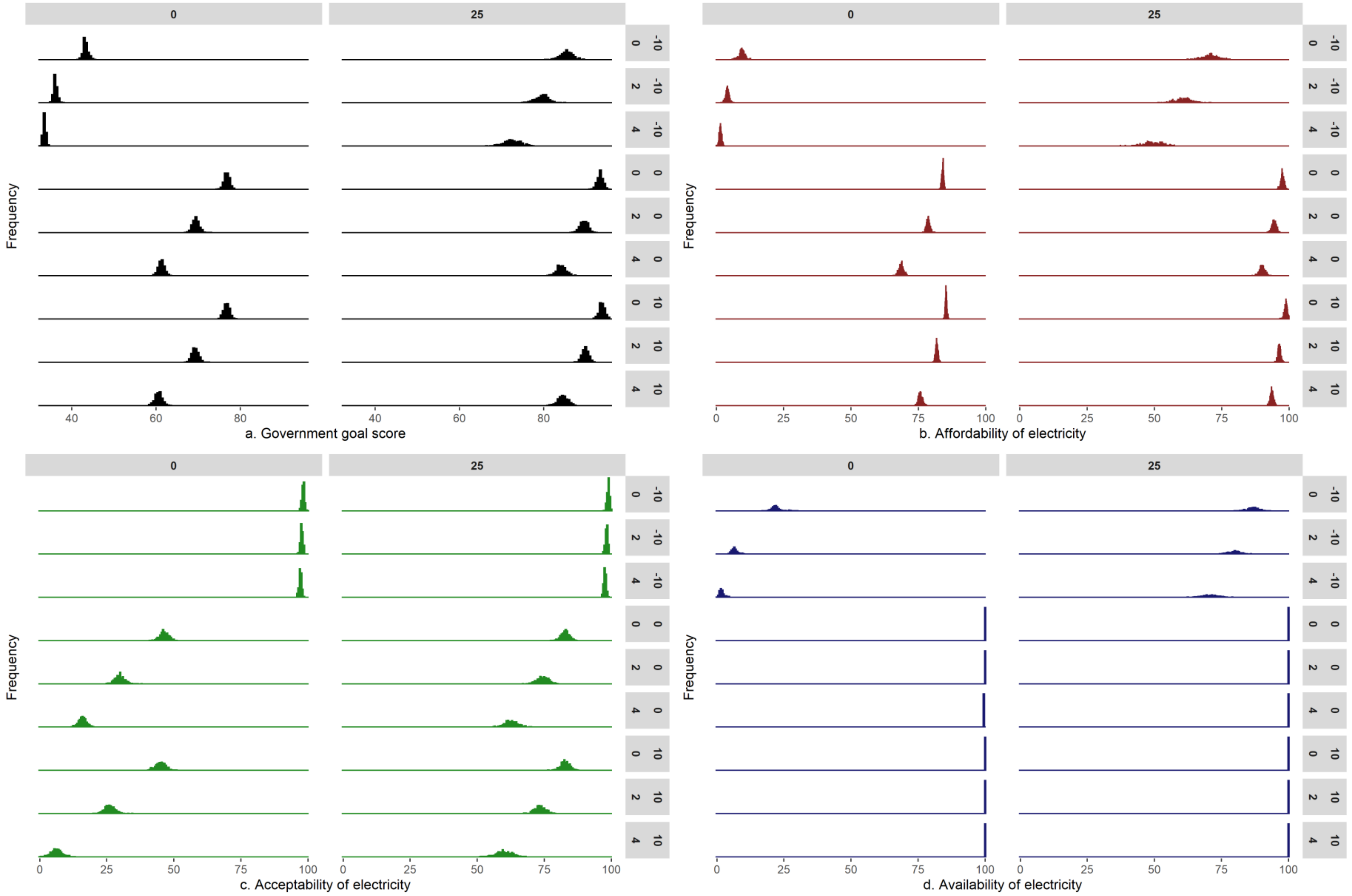

*Figure 6 Histogram of government goal score and its elements with respect to 1. Annual RES growth rate (0,25)%, 2. Annual non-RES growth rate (-10,0,10)%, and 3. Annual demand growth rate (0,2,4)%| Generally, more RES and non-RES growth results in better score for government goals; and higher demand growth decreases the value of government goal score*

# 4 Discussion

In our previous work[7], we named four areas of uncertainty for designing business models for ESS. The areas were 1) rules and regulations, 2) technology, 3) fiscal framework, and 4) market uncertainties. The results of experiments provide some additional insights into these four areas of uncertainty:

- **At least in the extreme conditions, carbon pricing is irrelevant.** With regard to the aspect of rules and regulations, in this analysis, we explored the opportunities for ESS under the current regulatory framework. For example, we limited ownership of grid-scale ESS to energy companies. The only regulatory aspect in the current analysis was the decision on the growth rate of carbon price. In fact, we considered a kind of $CO_2$ tax for generated electricity. However, interestingly, we found no significance for $CO_2$ price growth for the goals of the government and profitability of ESS, at least in the extreme conditions (e.g. extremely high or low RES, non-RES, etc.) of our experiments. It is noteworthy that by assigning the ownership of EES in the context of the market to energy companies we could model the interplay between storing capacity and, RES and non-RES, generation capacity. In our model it was thus possible to evaluate how different conditions favored or hindered the decisions of energy companies to develop their storing capacity in a context of growing carbon prices (tax) and how those fed back on that profitability.
- **Market penetration is still hanging on R&D and cost reduction.** We modeled ESS with certain technical and economic parameters which lead to negative average NPV for the majority of the experiment scenarios (More precisely, 87% of the scenarios in our experiments lead to a negative NPV for ESS, which can also be observed in Figure 5). Given the assumptions of the model, we found that increasing the energy capacity of ESS as well as reducing costs of the ESS energy unit is a key for profitability of ESS. We also found that a cheap ESS energy unit may cancel out the effects of an expensive ESS power unit in the long run. These insights explain why some electricity storage technologies such as the pumped hydro (with high energy and power rating, and relatively low energy capital cost) have worked better in the current systems. It also explains why other ESS technologies with the capability of scaling up to larger units (via stacking, etc.) find potential if their costs will decrease. In addition, improving the round-trip efficiency is another key for profitability of ESS, and it should be on the agenda of ESS developers.
- **At least in the extreme conditions, the wholesale market alone does not provide a solid justification for EES as an asset to pursue public policies aiming at stability and sustainability.** Although the fiscal framework in our model and analysis was omitted, some useful insights can still be drawn to provide recommendations for policy makers. We found that within the assumptions of our model and in the extreme conditions explored, ESS does not contribute significantly to the environmental and economic societal goals. Therefore, the tangible benefits of ESS for the government should be justified by ESS developers. If the societal aspect of ESS can be justified, the subsidies to decrease the capital costs of the energy unit will be recommended.

- **Deployment of EES does not reduce profitability, on the contrary, a high share of RES does so.** We explored market uncertainties by including two markets and their underlying mechanisms in our modeling practice. Adoption of agent-based modeling helped us to incorporate market dynamics and its randomness in our analysis. There is a hypothesis that the larger deployment of ESS may result in less profitability for ESS as it narrows the price gaps, however our results showed that higher total ESS power and energy capacity may result in higher profits. In addition, it is sometimes mentioned that ESS will find a business case if at least certain share of electricity will be generated from RES (e.g. [35] explains that variability of renewables can be accommodated by the grid itself when up to 20% of electricity is from RES and can be accommodated by other flexibility option when up to 35% of electricity generation is from RES). However, we found that a very high share of RES may reduce profitability of ESS. Therefore, we expect ESS to make profit within a range of RES capacity scenarios with both a minimum and a maximum.

Although our findings and results can shed a light into the fate of EES and mitigate some of the uncertainties of concern, this practice of modeling and analysis come with some limitations due to the simplifying assumptions, data unavailability, and computational limitations. Regarding the assumptions, we included only two single-application business models for ESS in this analysis. In our previous work [6] we provided a map of single-application business models and we emphasized the value of 'benefit stacking'. However, in this analysis, other business models and benefit stacking were absent. Although at the moment most emphasis is on the potential profitability of ESS for behind-the-meter or ancillary service applications (as we already discussed in [7]), and these business models were not covered in this analysis, we believe the current analysis forms a good basis to analyze more business models and their combinations in the future. In addition, we excluded cross-border trading, investment dynamics and carbon market dynamics from the model, and we assumed each month is represented by one day. Furthermore, the growth rate of RES was considered as an exogenous variable, thus leaving out limiting factors affecting productivity, such as climate variability or land use. The analysis may lead to different results by inclusion of such factors and mechanisms; it could enable new feedback loops in the actors presently unattainable. Moreover, because of unavailability of data, the value of some parameters such as the consumers' willingness to pay were determined by model calibration. Including real-world data for such parameters may open new windows and provide additional insights. In addition, we employed NetLogo as an open-source tool for rapid development of the model. However, in large settings and experiments, NetLogo models are slower than their counterparts in other programming environments in which data structures are determined by the modeler. In addition, NetLogo has a limitation for the number of experiments which forced us to include less values for the experiment parameters, and to conduct some parts of the experiments separately. Besides, although NetLogo experiments can be conducted with multiple cores of a single computer, we could not run the NetLogo experiments on multiple nodes of a computer cluster yet.

Last but not least, we tried to test business models in this analysis. However, testing business models in practice is a process and this quantitative research is just the first step to know the potential.

# 5 *Conclusion and Policy Implications*

In this paper we presented details of design and implementation of our exploratory agent-based modeling analysis approach to explore and test two single-application ESS business models in the context of the Dutch electricity market under deep uncertainties.

The objective of this paper was to address the question of the effects of grid-scale ESS business models on the investor's and the societal interests under deep uncertainties. Furthermore, the adoption of exploratory modeling analysis approach unraveled under what conditions ESS business models lead to benefits for the investors and the society. We found that investors would have more chance for profit by selecting the "wholesale arbitrage" business model, and we found that the profit depends more on the economics and technical characteristics of ESS (such as size and cost of energy ESS energy unit) than market and regulatory factors. In addition, we found no evidence in our results about the value of ESS for societal goals under extreme conditions explored. Besides, we found that low capital cost for ESS energy units and high round-trip efficiency are keys for profitability of ESS.

In our future work, we will incorporate investment dynamics and benefit stacking. The former enables us to study the co-evolution of ESS and the electricity market, and the latter may lead to more profitable spots.

# Acknowledgement

This research was funded by University of Groningen project code 190152210. In addition, we would like to thank the Center for Information Technology of the University of Groningen for their support and for providing access to the Peregrine high performance computing cluster.

# Appendix A. Model elements

*Table 5-1 Agent's attributes, properties, decisions and behaviors*

| Agents | Attributes and properties | Main decisions and behaviors |
|---|---|---|
| Energy Companies (producers and retailers) | - controls: power plant / ESS<br>or<br>- represents: small loads<br>----------------------------<br>- Type : producer or retailer<br>- ESS desirability<br>- Supply plan<br>- Demand plan<br>- Balancing plan<br>- Bank balance<br>- Surplus | - Investment in ESS and its business models<br>- Production Planning / Demand Forecasting<br>- Bidding/Offering in the wholesale market<br>- Bidding/Offering in the balancing market<br>- Calculating surplus<br>- manage electricity generation<br>- manage balancing energy provision<br>- Paying/Earning money<br>- managing bank account |
| Large-Consumers | - Owns large loads<br>----------------------------<br>- Demand plan<br>- Balancing plan<br>- Bank Balance | - Demand Forecasting<br>- Offering in the wholesale market<br>- Offering in the balancing market<br>- Paying/Earning money<br>- Calculating surplus |
| | - Surplus | - Manage balancing energy provision |
| Market | Controls: grid<br>----------------------------<br>- Electricity program<br>- Upward balancing bid ladder<br>- Downward balancing bid ladder<br>- Balancing program<br>- Curtail data<br>- CO2 revenue<br>----------------------------<br>Shared with the environment:<br>- Wholesale Supply Curve<br>- Wholesale Demand Curve<br>- Wholesale volume<br>- Wholesale price<br>- Power shortage size<br>- Power excess size<br>- Balancing volume<br>- Balancing price<br>- Balancing direction<br>- Blackout counter | - Collecting bids and offers<br>- Processing bids and offers<br>- Clearing markets<br>- Communicating e-programs<br>- Communicating balancing programs<br>- Calculate balancing fines<br>- Managing $CO_2$ payments<br>- Managing blackouts<br>- Collecting and distributing cash |

*Table 5-2Physical entities' attributes and behaviors*

| Physical Entities | Attributes and properties | Behaviors |
|---|---|---|
| Power Plants | - Flexible [Boolean]<br>- Available [Boolean]<br>- Marginal cost (€/MWh) [float]<br>- Capacity (MW) [integer]<br>- Fuel [String]<br>- Technology [String]<br>- Reliability (%)<br>- Efficiency (%)<br>- Generation (MWh)<br>- emission (Ton $CO_2$)<br>- capital cost (€/MW)<br>- fixed O&M (€/MW/y)<br>- variable O&M (€/MW/y) | - Inject to grid<br>- emit $CO_2$ |
| ESS | - Flexible [Boolean]<br>- Available [Boolean]<br>- Energy Capacity (MWh)<br>- Power Capacity (MW)<br>- Marginal cost (€/MWh)<br>- Content (MWh)<br>- Generation (MWh)<br>- Consumption (MWh)<br>- Purchase cost (€/MWh)<br>- Revenue (€)<br>- NPV (€)<br>- Business Model [String] | - Withdraw from grid<br>- Inject to grid<br>- Update content<br>- Update marginal cost |
| Large Loads | - Flexible [Boolean]<br>- Yearly consumption (MWh)<br>- Energy needs (MWh)<br>- Willingness to Pay (€/MWh)<br>- Consumption (MWh) | - Consume electricity<br>- Withdraw from grid |
| Small Loads | - Yearly consumption (MWh)<br>- Energy needs (MWh)<br>- Willingness to Pay (€/MWh)<br>- Consumption (MWh) | - Consume electricity<br>- Withdraw from grid |
| Grid | - Inflow (MWh)<br>- Outflow (MWh)<br>- Imbalance (MWh)<br>- Imbalance details [list] | - Check grid balance<br>- Communicate imbalance |

*Table 5-3 Parameters and variables of the environment*

| Domain | Type | Variables |
|---|---|---|
| ESS | Parameter | - ESS Business Models (wholesale arbitrage and reserve capacity)<br>- Desirability of ESS for energy companies (%)<br>- Grid ESS capacity (MW)<br>- Maximum ESS energy rating (MWh)<br>- ESS power capital cost (€/MW)<br>- ESS energy capital cost (€/MWh)<br>- ESS fixed O&M costs (€/MW/y)<br>- ESS roundtrip efficiency (%) |
| Regulation | Parameter | - $CO_2$ pricing [Boolean]<br>- $CO_2$ price: Initial value and growth rate |
| Generation | Parameter | - Number of energy producers<br>- Total generation capacity (GW)<br><br>Generation Technologies:<br>- Initial RES (%)<br>- share (% of non-RES) for nuclear, coal, CCGT, OCGT<br>- share (% of RES) for Offshore/onshore wind, solar<br>- Annual growth rate (%/y) for RES and non-RES<br><br>Generation Inputs:<br>- Wind availability time series<br>- Solar availability time series<br>- fuel prices patterns<br>- fuel prices: Initial value (€/unit) and growth rate (%) |
| Demand | Parameter | - Number of large consumers<br>- Total large consumer demand: Initial (GW/y) and growth rate (%/y)<br>- Percentage of flexible loads (%)<br>- Large consumer willingness-to-pay: Minimum (€/MWh) and range (€/MWh)<br><br>- Number of retailer energy companies<br>- Total small consumer demand: Initial (GW/y) and growth rate (%/y)<br>- Small consumer willingness-to-pay: Minimum (€/MWh) and range (€/MWh)<br>- Small consumers load profile |
| General | variables | - Interest rate (%)<br>- length of memory (Month)<br>- Hour Type (peak/off-peak) |
| | | - Electricity price history<br>- wind condition history<br>- wind forecast<br>- Total CO2 Emission<br>- CO2 price: current and history |

## Appendix B. Model input data

The model gets some inputs. In addition to the data determined by scenarios of in the experiments (see Table 1), some input data are shared in all experiments and they are either derived from literature, or achieved during model calibration process, and the rest are set according to scenarios. Table 5-1 illustrates the main data inputs of the model, sub-elements of data, their value range, and the references from which data are derived if applicable.

*Table 5-1 Model inputs and assumptions*

| Input | Unit | Sub-elements | Value range | Source |
|---|---|---|---|---|
| Number of agents | # | Energy producing Co.<br>Retailer Co.<br>Large Consumer Co. | 5<br>8<br>16 | [6]<br>[6]<br>Calibration |
| Baseline time series of coal prices | €/ton | | 53.97-132.37 | [36] |
| Baseline time series of natural gas prices | €/MWh | | 13.68-2363 | [37] |
| Baseline time series of Uranium prices | €/kg | | 43.55-95.56 | [38] |
| Baseline time series of wind availability | % | | 47-100 | [39] |
| Baseline time series of sun availability | % | | 0-100 | [39] |
| Baseline time series of hourly load profiles | % | E1a | 0.0053-0.0229 | [40,41] |
| Total load | GWh/y | | 100,000 | [42] |
| Total Generation Capacity | GW | | 31.1 | [43] |
| Generation technologies- Initial share (in 2016) | % | Nuclear<br>Coal<br>Open-cycle gas<br>Combined-cycle gas<br>Wind-offshore<br>Wind-onshore<br>Solar | 1.6<br>18.2<br>32<br>32<br>1.1<br>10.5<br>4.6 | [43] |
| Generation technologies- Growth rate (2016-2018) | %/y | Nuclear<br>Coal<br>Open-cycle gas<br>Combined-cycle gas<br>Wind-offshore<br>Wind-onshore | 0<br>-12<br>0<br>0<br>65<br>35 | [43] |
| | | Solar | 5 | |
| Generation technologies- Thermal Efficiency | % | Nuclear | 33 | [44] |
| | | Coal | 40 | |
| | | Open-cycle gas | 38.5 | |
| | | Combined-cycle gas | 56 | |
| | | Wind-offshore | - | |
| | | Wind-onshore | - | |
| | | Solar | - | |
| Generation technologies- Reliability | % | Nuclear | 85 | |
| | | Coal | 86 | |
| | | Open-cycle gas | 80 | |
| | | Combined-cycle gas | 84 | |
| | | Wind-offshore | 95 | |
| | | Wind-onshore | 95 | |
| | | Solar | 99 | |
| Consumers' maximum willingness to pay | €/MWh | Large consumers | 0-150 | Calibration |
| | | Small consumers | 200 | |

Variables such as ESS charge/discharge duration, operation and maintenance cost, and self-discharge rate are not incorporated in the model or experiments at this stage as a simplifying assumption.

# Appendix C. Bidding Strategies

In this section, we provide detailed formulas for calculation of bidding strategies of wholesale market participants.

## C-1. Power Plants

Offered Energy:

$$Offered\ Energy\ per\ Hour\ [MWh] = Power\ Capacity$$

Without Carbon Pricing:

$$Marginal\ Cost = \frac{Fuel\ Price\ [\frac{Euro}{Wright\ Unit}]}{Efficiency * Energy\ Value[\frac{MWh}{Weight\ Unit}]} + Variable\ O\&M\ Cost[\frac{Euro}{MWh}]$$

With Carbon Pricing:

$$Marginal\ Cost = \frac{Fuel\ Price\ [\frac{Euro}{Wright\ Unit}]}{Efficiency * Energy\ Value[\frac{MWh}{Weight\ Unit}]} + Variable\ O\&M\ Cost[\frac{Euro}{MWh}] + \frac{Fuel\ Carbon\ Content\ \left[\frac{Ton\ CO_2}{MWh}\right]}{Efficiency} * CO_2\ Price[\frac{Euro}{Ton\ CO_2}]$$

Offer Structure:

$$Offer = \{Offered\ Energy, Marginal\ Cost\}$$

## C-2. Loads

Demanded Energy:

$$Demanded\ Energy\ per\ Hour\ [MWh] = Power\ Capacity$$

Bid Structure:

$$Bid = \{Demanded\ Energy, Maximum\ Willingness\ to\ pay\}$$

## C-3. ESS

**Off-peak Hours:**

$$Charge\ Potential[MWh] = Min(Power\ Capacity[MW], Energy\ Capacity[MWh] - Energy\ Content[MWh])$$

$$Willingness\ To\ Pay = \sum_{t=1}^{24} \frac{Electricity\ Price_{t-1}}{24}$$

$$Bid = \{Charge\ Potential, Willingness\ To\ Pay\}$$

**Peak Hours:**

$$Discharge\ Potential[MWh] = Min(Power\ Capacity[MW], Energy\ Content[MWh])$$

$$Marginal\ Cost = \frac{\sum Charged\ Energy(t) * Purchase\ Price\ (t)}{Energy\ Content}$$

$$Offer = \{Discharge\ Potential, Marginal\ Cost\}$$

# Appendix D. Model processes sequence

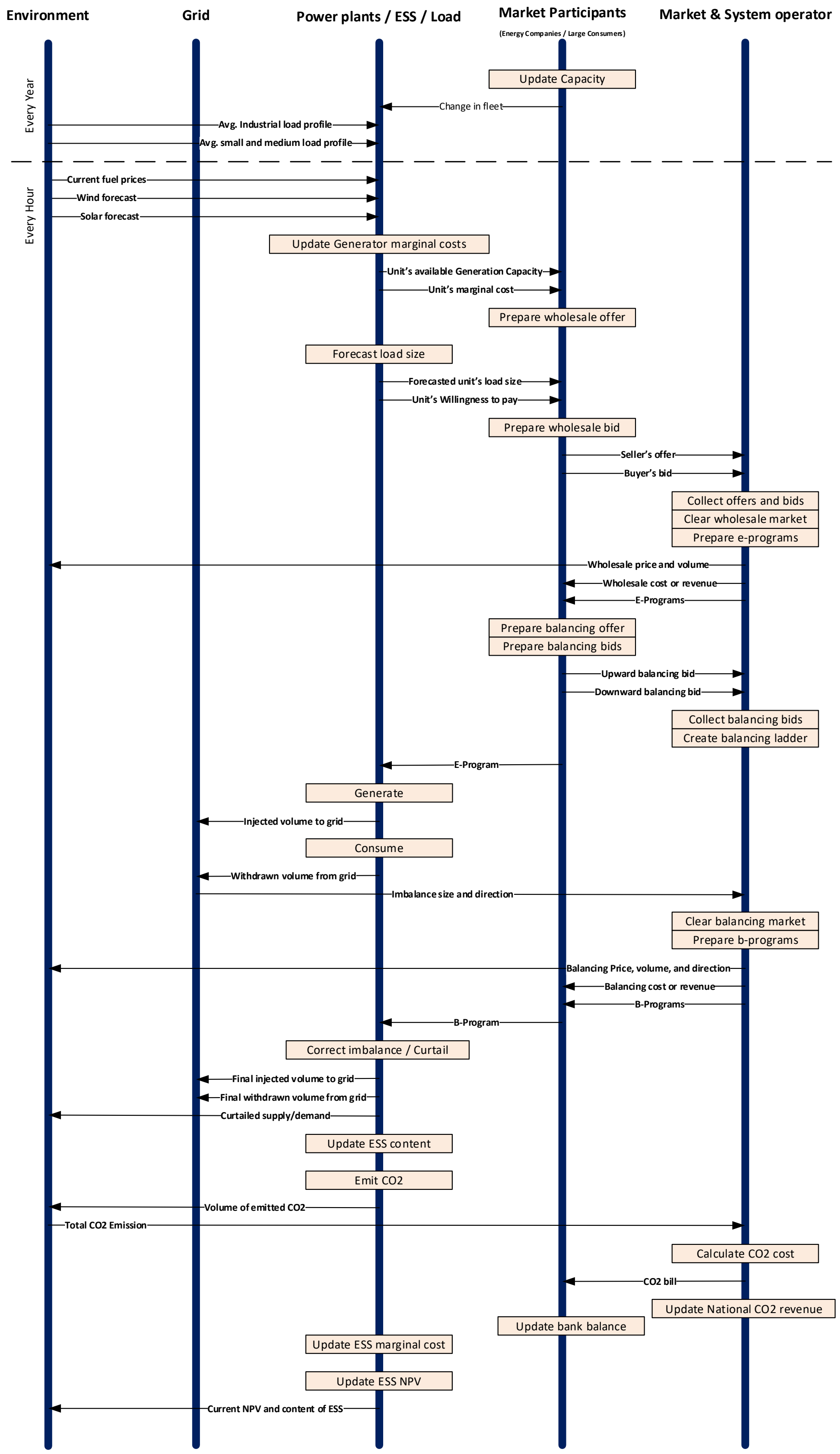

*Figure 5-1 Sequence of model processes*

## Appendix E. Pseudo Code

In this section we provide pseudo code of the model algorithm and some sub-model algorithms. The main model has two procedures of setup and main. The **bolded** words indicate the standard flow of the code. After the word **do**, the model elements which are called before "**do**" run the sub-algorithms between two braces {}. The Capitalized words refer to the environment, agents, or physical entities, or their properties. Phrased preceded by # are comments and additional information.

### Model

```
procedure setup
        create agents                     # - create Market, Energy Producing Companies,
                                          # Energy Retailer Companies, and Large Consumers
        create physical entities          # - create Power Plants, Large Loads, Small Loads,
                                          # ESS, and Transmission Grid
        create links                      # - create Information, Cash Flow, Control, Supply
                                          # Contract, and Wire links among agents and/or
                                          # physical entities
end procedure
procedure main
while simulation time < Time Limit Do
        update generation capacity                              # add or remove power plants
                                                                # based on growth rate parameters
        Agents do { take roles}                                 # set "Large Producer?", "Retailer?",
                                                                # "Large Consumer?", and "Market"
        Large Producers do {calculate producer wholesale bids}  # quantity = available capacity
                                                                # price = marginal cost
        Large Consumers do {calculate consumer wholesale bid}
        Retailers do {calculate retailer wholesale bid}
        Large Producers, Retailers, and Large Consumers do {
                send  producer wholesale bids to Information link}
        Markets  do {
                collect wholesale bids from Information links
                clear wholesale market                          # determine market price and
                                                                # volume
                calculate wholesale e-programs
                send wholesale contracts to information links}
        Large Producers, Large Consumers, and Retailers do {
                deposit/withdraw wholesale money in/from Cash Flow links
                update bank balance}
        Large Producers do {
                if controlling flexible fleet  then
                        calculate producer balancing bid
                        send producer balancing bid to Information links
                end if}
        Large Consumer do {
                if controlling flexible fleet then
                        calculate Large Consumer balancing bid
                        send Large Consumer balancing bid to Information links
                end if}
```

```
Markets  do {
        collect balancing bids from Information links
        create balancing bid ladder}                    # including all bids for upward and
                                                        #downward balancing energy
Loads do {
        consume electricity
        withdraw electricity from incoming Wire}
Large Producers do {
        if controlling ESS then
                order controlled ESS to do {
                        Charge according to e-program
                        withdraw electricity from  incoming Wire}
        end if }
Large Producers do {
        Order controlled power plants or ESS in the e-program to do {
                generate electricity
                inject electricity to power plant's or ESS's outgoing Wire}
                }
Transmission Grids do {
        check electricity balance
        if any imbalance  then
                send imbalance volume and direction to Market via Control Link
        end if }
if any imbalance then
        Markets do {
                clear balancing market                  # determine market price,
                                                        # volume and direction
                prepare balancing Programs
                send balancing contracts to other agents via Information links}
end if
Energy Producing Companies do {
        if balancing contract received then provide producer balancing Energy end if }
Large End Consumers do {
        if balancing contract received then provide consumer balancing Energy end if }
markets do {
        if balancing market fails to solve imbalance then
                if excess of supply then curtail end if      #unplug some generation
                if deficit of supply then blackout end if  #update blackout variable
                withdraw imbalance fine from Cash Flow links with balance deviators
                deposit balancing revenue to Cash Flow links with service providers
        end if }
Power Plants do {
        if fuel = "Natural Gas" OR "Coal" then emitCO2 end if }
Markets do {
        calculate Total CO2 Emission
        if Carbon Pricing? then
                withdraw CO2 money from Cash Flow links with CO2 emitting agents
        end if}
Large Producers, Large Consumers, and Retailer do {
        update Bank Balance}                            # based on Cash Flow links
ESSs do {
```

```
        Update ESS status}                          # update marginal cost and NPV
    Update Environment variables                    # price history, CO2 price, etc.
    Simulation time = simulation time + 1
end while
End procedure
```

## Calculate producer wholesale bids

```
procedure calculate producer wholesale bids         # run by Energy Producing Company
make Bid List empty
for all owned Power Plants do {
    if technology = "Wind-Offshore" OR "Wind-Onshore" OR "Solar-PV" then
        if technology = "Solar-PV" then
            Bid Volume = capacity * solar availability
        else
            Bid Volume = capacity * wind availability
        end if
        Marginal Cost = 0
    else
        Bid Volume = Power Plant's capacity
        if carbon pricing  then
            if fuel = "Natural Gas" OR "Coal" then
                Marginal Cost = fuel price / efficiency / fuel energy value +
                                (fuel carbon content  / efficiency) *
                                estimated carbon price
            else
                Marginal Cost = fuel price / efficiency / fuel energy value
            end if
        else
            Marginal Cost = fuel price / efficiency / fuel energy value
        end if
    end if
    append (Bid Volume, Marginal Cost, "Supply") to Bid List
}
for all owned ESS do {
    if time = "Peak" then
        if ESS's Business Model = "Wholesale Arbitrage" then
            Bid Volume = Minimum (power capacity, content)
            append (Bid Volume, Marginal Cost, "Supply") to  Bid List
        end if
    else                                            # Off-Peak
        if ESS's Business Model = "Wholesale Arbitrage" OR "Reserve Capacity" then
            Bid Volume = Minimum (power capacity, energy capacity - content)
            append (Bid Volume, estimated electricity price, "Demand") to  Bid List
        end if
    end if
}
end procedure
```

## Calculate producer balancing bid

```
procedure calculate producer balancing bid          # run by Energy Producing Company
```

```
make Bid List empty
for all owned flexible Power Plants do {
        if power plant is involved in the current wholesale e-program then
                if e-program volume = power plant capacity then
                        Bid Volume = power plant capacity
                        append (Bid Volume, - Marginal Cost, “Downward”) to Bid List
                else
                        Bid Volume = power plant capacity – contracted generation in e-program
                        append (Bid Volume, Marginal Cost, “Upward”) to Bid List
                end if
        else
                Bid Volume = power plant capacity
                append (Bid Volume, Marginal Cost, “Upward”) to Bid List
        end if
}
for all owned ESS do {
        if Business Model = “Reserve Capacity” then
                Bid Volume = Minimum (power capacity, content)
                append (Bid Volume, Marginal Cost, “Upward”) to Bid List
        end if
}
end procedure
```

## Update ESS status

```
procedure update ESS status                                   # run by ESS
        if business model = "Wholesale Arbitrage" then
                if off-peak hour then
                        Marginal Cost = (content t-1*Marginal Cost t-1+
                                        Charged Volume t * Wholesale price t) / content t
                        Purchase Cost =  Charged Volume * Wholesale price
                        NPV = NPV - 30 * Purchase Cost / (1 + interest rate /100)^year
                else
                        Revenue = Discharged Volume * Wholesale price
                        NPV = NPV + 30 * Revenue  / (1 + interest rate /100)^year

                end if
                if beginning of the year then
                        NPV = NPV – (O&M Cost * capacity ) / (1 + interest rate / 100)^year
                end if
        end if
        if business model = "Reserve Capacity" then
                Revenue= Discharged Volume * Balancing price
                NPV = NPV + 30 * Revenue / (1 + interest rate / 100)^year
                If off-peak hour then
                        Marginal Cost = (content t-1*Marginal Cost t-1+
                                        Charged Volume t * Wholesale price t) / content t
                        Purchase Cost =  Charged Volume * Wholesale price
                        NPV = NPV –  30 * Purchase Cost / (1 + interest rate /100)^year
                end if
                if beginning of the year then
                        NPV = NPV – (O&M Cost * capacity ) / (1 + interest rate / 100)^year
```

**end if**

**end if**

**end procedure**

# Appendix F. Evaluation Criteria Calculation

## F-1. Profitability

$$Capital\ cost_k = \text{Power capacity}_k * \text{ESS power capital cost} + \text{energy capacity}_k * \text{ESS energy capital cost}$$

$$NPV_k = Capital\ Cost_k + \sum_{t=1}^{5760} \frac{(Revenue_k(t) - Purchase\ Cost_k(t) - \frac{Fixed\ O\&M\ Cost}{288})}{\left(1 + \frac{i}{100}\right)^{\left(\left\lfloor\frac{t-1}{288}\right\rfloor + 1\right)}}$$

$$RunNPV_r = \sum_{k=1}^{n} \frac{NPV_k}{n}$$

$$ScenarioNPV_S = \sum_{r=1}^{20} \frac{RunNPV_r}{20}$$

$$Profitability_s = \frac{ScenarioNPV_S - Min(ScenarioNPV)}{Max(ScenarioNPV) - Min(ScenarioNPV)} * 100$$

## F-2. Affordability

$$Run\ Price_r = \sum_{t=1}^{5760} \frac{Day - ahead\ Price\ (t)}{5760}$$

$$Scenario\ Price_s = \sum_{r=1}^{20} \frac{Run\ Price_r}{20}$$

$$Affordability = \frac{Max(Scenario\ Price) - Scenario\ Price_s}{Max(Scenario\ Price) - Min(Scenario\ Price)} * 100$$

## F-3. Availability

$$Total\ Run\ Blackout_r = Blackout\ counter(5760)$$

$$Scenario\ Blackout_s = \sum_{r=1}^{20} \frac{Total\ Run\ Blackout_r}{20}$$

$$Availability_s = \frac{Max(Scenario\ Blackout) - Scenario\ Blackout_s}{Max(Scenario\ Blackout) - Min(Scenario\ Blackout)} * 100$$

## F-4. Acceptability

$$Total\ Run\ Emission_r = Total\ CO_2\ Emission\ (5760)$$

$$Scenario\ Emission_s = \sum_{r=1}^{20} \frac{Total\ Run\ Emission_r}{20}$$

$$Acceptability = \frac{Max(Scenario\ Emission) - \ Scenario\ Emission_s}{Max(Scenario\ Emission) - \ Min(Scenario\ Emission)} * 100$$